\documentclass[11pt]{article}
\usepackage{graphicx}

\newcommand{\BABARPubYear}    {00}

\newcommand{\BABARProcNumber} {34}
\newcommand{\SLACPubNumber} {8748}

\def\lbabar{\mbox{{\large\sl B}\hspace{-0.4em} {\normalsize\sl A}\hspace{-0.03em}{\large\sl B}\hspace{-0.4em} {\normalsize\sl A\hspace{-0.02em}R}}}
\usepackage{relsize}
\def\babar{\mbox{\slshape B\kern-0.1em{\smaller A}\kern-0.1em
    B\kern-0.1em{\smaller A\kern-0.2em R}}}

\def\epem       {\ensuremath{e^+e^-}}

\def\mumu       {\ensuremath{\mu^+\mu^-}}

\def\qqbar {\ensuremath{q\overline q}}

\def\piz   {\ensuremath{\pi^0}}

\def\pip   {\ensuremath{\pi^+}}
\def\pim   {\ensuremath{\pi^-}}
\def\pipi  {\ensuremath{\pi^+\pi^-}}

\def\Kbar  {\kern 0.2em\overline{\kern -0.2em K}{}}

\def\Kp    {\ensuremath{K^+}}
\def\Km    {\ensuremath{K^-}}
\def\KS    {\ensuremath{K^0_{\scriptscriptstyle S}}} 
 
\def\Kstarz  {\ensuremath{K^{*0}}}

\def\Kstar   {\ensuremath{K^*}}

\def\Kstarp   {\ensuremath{K^{*+}}}

\def\Kzb   {\ensuremath{\Kbar^0}}
\def\KzKzb {\ensuremath{K^0 \kern -0.16em \Kzb}}
\def\Dz    {\ensuremath{D^0}}
\def\Dbar  {\kern 0.2em\overline{\kern -0.2em D}{}}

\def\Dzb   {\ensuremath{\Dbar^0}}
\def\DzDzb {\ensuremath{D^0 {\kern -0.16em \Dzb}}}

\def\Bz    {\ensuremath{B^0}}
\def\B     {\ensuremath{B}}
\def\Bbar  {\kern 0.18em\overline{\kern -0.18em B}{}}

\def\Bzb   {\ensuremath{\Bbar^0}}
\def\Bu    {\ensuremath{B^+}}

\def\BB    {\ensuremath{B\Bbar}} 
\def\BzBzb {\ensuremath{B^0 {\kern -0.16em \Bzb}}}

\def\jpsi  {\ensuremath{{J\mskip -3mu/\mskip -2mu\psi\mskip 2mu}}} 
\def\psitwos {\ensuremath{\psi{(2S)}}}
\mathchardef\Upsilon="7107
\def\Y#1S{\ensuremath{\Upsilon{(#1S)}}}

\def\FourS {\Y4S}

\mathchardef\Deltares="7101
\mathchardef\Xi="7104
\mathchardef\Lambda="7103
\mathchardef\Sigma="7106
\mathchardef\Omega="710A
\def\Deltabar   {\kern 0.25em\overline{\kern -0.25em \Deltares}{}}
\def\Lbar {\kern 0.2em\overline{\kern -0.2em\Lambda\kern 0.05em}\kern-0.05em{}}
\def\Sigbar{\kern 0.2em\overline{\kern -0.2em \Sigma}{}}
\def\Xibar{\kern 0.2em\overline{\kern -0.2em \Xi}{}}
\def\Obar{\kern 0.2em\overline{\kern -0.2em \Omega}{}}
\def\Nbar{\kern 0.2em\overline{\kern -0.2em N}{}}
\def\Xbar{\kern 0.2em\overline{\kern -0.2em X}{}}

\def\BR{{\ensuremath{\cal B}}}

\def\mes        {\mbox{$m_{\rm ES}$}}

\def\ev   {\ensuremath{\rm \,e\kern -0.08em V}}
\def\kev  {\ensuremath{\rm \,ke\kern -0.08em V}} 
\def\mev  {\ensuremath{\rm \,Me\kern -0.08em V}} 
\def\gev  {\ensuremath{\rm \,Ge\kern -0.08em V}} 
\def\gevc {\ensuremath{{\rm \,Ge\kern -0.08em V\!/}c}} 
\def\tev  {\ensuremath{\rm \,Te\kern -0.08em V}}
\def\mevc {\ensuremath{{\rm \,Me\kern -0.08em V\!/}c}} 
\def\gevcc{\ensuremath{{\rm \,Ge\kern -0.08em V\!/}c^2}} 
\def\mevcc{\ensuremath{{\rm \,Me\kern -0.08em V\!/}c^2}}

\def\mus  {\ensuremath{\rm \,\mus}}

\def\ps   {\ensuremath{\rm \,ps}}

\def\mus        {\ensuremath{\,\mu{\rm s}}}    
\def\ps         {\ensuremath{{\rm \,ps}}}   
\def\gsim{{~\raise.15em\hbox{$>$}\kern-.85em
          \lower.35em\hbox{$\sim$}~}}
\def\lsim{{~\raise.15em\hbox{$<$}\kern-.85em
          \lower.35em\hbox{$\sim$}~}}

\def\CP                 {\ensuremath{C\!P}}

\def\to                 {\ensuremath{\rightarrow}}

\def\pep2{PEP-II}
\def\BF{$B$ Factory}

\def\chic#1{\ensuremath{\chi_{c#1}}} 

\def\stwob{\ensuremath{\sin\! 2 \beta   }}

\def\mistag{\ensuremath{w}}

\def\deltaz{\ensuremath{{\rm \Delta}z}}
\def\deltat{\ensuremath{{\rm \Delta}t}}
\def\deltamd{\ensuremath{{\rm \Delta}m_d}}

\providecommand{\eqref}[1]{Eq.~(\ref{eq:#1})}

\newcommand{\epjc}      [1]  {{Eur.\ Phys.\ Jour.\ C~{\bf #1}}}

\providecommand{\pl}        [1]  {{Phys.\ Lett.\ {\bf #1}}}      

\providecommand{\prl}       [1]  {{Phys.\ Rev.\ Lett.\ {\bf #1}}} 
\providecommand{\pr}        [1]  {{Phys.\ Rev.\ {\bf #1}}}

\def\jetset74   {\mbox{\tt Jetset \hspace{-0.5em}7.\hspace{-0.2em}4}}

\newcommand{\BaBar}{\textsc{BaBar}}

\newcommand{\BNotJKz}{\Bz$\rightarrow$\jpsi$K^0$}
\newcommand{\BpJKp}{\Bu$\rightarrow$\jpsi\Kp}

\newcommand{\BNotJKstarz}{\Bz$\rightarrow$\jpsi\Kstarz}

\newcommand{\BpJKstarp}{\Bu$\rightarrow$\jpsi\Kstarp}
\newcommand{\BNotpsitwosKs} {\Bz$\rightarrow$\psitwos\KS}
\newcommand{\BNotpsitwosKz} {\Bz$\rightarrow$\psitwos$K^0$}
\newcommand{\BppsitwosKp} {\Bu$\rightarrow$\psitwos\Kp}

\newcommand{\BpChiKp} {\Bu$\rightarrow$\chicone\Kp}
\newcommand{\toLL} {$\rightarrow \ell^+\ell^-$}

\newcommand{\DE}{\mbox{$\Delta E$}}

\def\pipi     {\ensuremath{\pi\pi}}
\def\kpi     {\ensuremath{K\pi}}
\def\kk     {\ensuremath{KK}}
\def\fish    {\ensuremath{\cal F}}
\def\psitwos {\ensuremath{{\psi(2S)}}}
\def\chic {\ensuremath{{\chi_c}}}
\def\chicone {\ensuremath{{\chi_{c1}}}}
\def\chictwo {\ensuremath{{\chi_{c2}}}}
\def\Acp {\ensuremath{{\cal A}_{CP}}}
\def\YfourS {\ensuremath{\Upsilon(4S)}}

\newcommand{\AmS}{{\protect\the\textfont2
  A\kern-.1667em\lower.5ex\hbox{M}\kern-.125emS}}

\long\def\inst#1{\par\nobreak\kern 4pt\nobreak
    {\it #1}\par\vskip 10pt plus 3pt minus 3pt}

\begin{document}
{\pagestyle{empty}

\begin{flushright}
SLAC-PUB-\SLACPubNumber \\
\babar-PROC-\BABARPubYear/\BABARProcNumber \\
September, 2000 \\
\end{flushright}

\par\vskip 4cm

\begin{center}
\Large \bf The first physics results from {\BaBar}.
\end{center}
\bigskip

\begin{center}
\large 
G. Sciolla\\
Stanford Linear Accelerator Center\\
 Mail Stop 95, P.O. Box 4339, Stanford, CA 94309, USA \\
(for the \lbabar\ Collaboration)
\end{center}
\bigskip \bigskip

\begin{center}
\large \bf Abstract
\end{center}
The {\BaBar} experiment and the PEP-II accelerator at SLAC started
to take data on May 26, 1999.
By the time of this conference, the recorded 
integrated luminosity was 20 fb$^{-1}$,
of which 8 fb$^{-1}$ were  analyzed to provide 
a first set of physics results.
This talk reviews
the first measurement of $\sin2\beta$ and 
the study of B meson decays to charmonium modes
and 2-body charmless decays. 
Complementary  results were presented by
the \BaBar\ Collaboration at this 
conference and are reviewed in~\cite{mhs}.  

\vfill
\begin{center}
Contribued to the Proceedings of the International Conference On CP Violation Physics,  \\
                            18-22 Sep 2000, Ferrara, Italy 
\end{center}

\vspace{1.0cm}
\begin{center}
{\em Stanford Linear Accelerator Center, Stanford University, 
Stanford, CA 94309} \\ \vspace{0.1cm}\hrule\vspace{0.1cm}
Work supported in part by Department of Energy contract DE-AC03-76SF00515.
\end{center}


\section{Physics at \babar }

The {\BaBar} experiment~\cite{tdr} is designed in order to perform
 a comprehensive  study of CP violation in the B system. 
The main goal is to investigate if CP violation can be fully 
explained within the Standard Model by the imaginary phase of the 
Cabibbo-Kobayashi-Maskawa matrix, 
or if other sources of CP violation should be looked for 
elsewhere~\cite{physbook}. 

The strategy is to over-constrain the Unitarity Triangle 
by measuring at the same time its sides and angles. 

The 
 sides can be extracted 
from measurements of $|V_{tb}^*V_{td}|$ in the study of 
$B^0 \overline B{^0}$ mixing, $V_{ub}$ from the study of 
decays of B mesons in charmless final states, and $V_{cb}$ 
from studies of the decay $B \to D^* l \nu$.
 
The determination of the angles requires to measure 
time dependent asymmetries for the decay of a neutral B meson into 
a CP eigenstate,  $f_{CP}$, of the form 
\begin{equation}
\Acp(t) =
{ \frac
{\Gamma ( B^0(t) \rightarrow f_{CP}) - \Gamma (\overline {B^0}(t) \rightarrow f_{CP}) } 
{\Gamma ( B^0(t) \rightarrow f_{CP}) + \Gamma (\overline {B^0}(t) \rightarrow f_{CP}) } 
} .
\label{eq:asym}
\end{equation}

For B$^0$ decays with only one diagram contributing to the final state 
\begin{equation}
\Acp(t) = - \mathrm{Im}\lambda \sin(\Delta m \Delta t),
\end{equation}
where $\lambda = p/q \overline A (f) / A(f)$, $A(f)$ and $\overline A (f)$ are the decay amplitudes 
and $p$ and $q$ are the complex coefficients that relate  mass and flavour eigenstates of the B mesons~\cite{physbook}. 
For some decay modes, Im$\lambda$ is directly and simply related to the angles
 of the Unitarity Triangle. Of particular interest are the decays of the 
neutral  B mesons in final states containing a charmonium state and 
a neutral kaon. For these decays
\begin{equation}
\Acp(t) = \pm\sin(2\beta ) \sin( \Delta m \Delta t).  
\end{equation} 

Experimentally, the measurement of the CP asymmetry $\Acp$
requires three basic steps: 
\begin{itemize}
\item the exclusive reconstruction of one B meson ($B_{CP}$)  
 in a CP eigenstate  (e.g. 
$B^0 \rightarrow J/\psi K_s$); 
\item the measurement of the time $\Delta t$ elapsed between the decay of the 
two neutral B mesons produced in the $\YfourS$ decay; 
\item the tagging of the CP eigenstate at the production time, inferred 
 from the tagging of the other B meson ($B_{tag}$) produced coherently in the  
$\YfourS$ decay.  
\end{itemize}



\section{The experimental apparatus} 

The PEP-II storage ring and the \BaBar\ detector have been
designed specifically to measure time-dependent CP asymmetries 
in the neutral B meson decays in the  most effective way.

\subsection{The PEP-II accelerator}

PEP-II is an asymmetric B factory 
consisting  of two concentric rings: a high energy ring (HER) with a 9.0 GeV 
electron beam and a low energy ring (LER) with a 3.1 GeV positron beam. 
The energy of the two beams  has been chosen
in order to achieve a center of mass 
energy of 10.58 GeV, the mass of the $\YfourS$ resonance,
and a boost of 0.56 of the $\YfourS$ along the $z$ axis.
The boost corresponds to an average separation of 260~$\mu$m
between the decay vertices of the two B mesons produced in
the $\YfourS$ decay.
The design luminosity of the machine is 3$\times 10^{33} s^{-1}cm^{-2}$, 
which corresponds 
to about 30 million \BB\ pairs produced per year. 

PEP-II started its activity in factory mode on May 26, 1999 and reached 
outstanding performance very rapidly.   
By the time of this conference the achieved peak luminosity 
was 2.56 $\times 10^{33} s^{-1}cm^{-2}$, which corresponds to 85\%
 of the design value. 
The design daily recorded luminosity of 135 pb$^{-1}$/day was routinely 
achieved  and exceeded starting in June 2000, with a record of 151 pb$^{-1}$ 
accumulated in one day.
By the time of this conference the total integrated luminosity
delivered by PEP-II 
was 19.5 fb$^{-1}$, of which 17.5 recorded at the $\YfourS$ resonance and 
2 fb$^{-1}$ at a center of mass about 40 MeV below resonance. 
Thanks to a data taking efficiency higher than 95\%,  \BaBar\  
accumulated  more than 18 fb$^{-1}$ on tape.
 
\subsection{The \babar\  detector}

\BaBar\ is a cylindrical detector built around the interaction 
point of PEP-II. It consists of several concentric subdetectors. 
From the inside out there are: 

\begin{itemize}
\item  a silicon vertex tracker, consisting of 5 layers of double-sided silicon  strips;   
\item  a cylindrical drift chamber, filled with a low density gas mixture (Helium-Isobutane: 80\%-20\%) to minimize multiple scattering; 
\item  the DIRC detector, the main particle identification device of the experiment, which consists of 144 quartz bars; 
\item  an electromagnetic calorimeter, consisting of 6580 CsI crystals; 
\item  a 1.5 Tesla superconducting magnet; 
\item  an instrumented flux return (IFR), consisting of 19 layers of RPCs, for muon identification and K$_L$ reconstruction.  
\end{itemize}

The combined tracking resolution  of the \BaBar\ experiment 
can be parameterized as 
\begin{equation}
\left({ \frac {\Delta p_T } { p_T } }\right)^2 = (0.0015 p_T)^2 + 0.005^2.  
\end{equation}

The performances of electron, muon and kaon identification are 
summarized in Table~\ref{tab:pid}.  
\begin{table}
\caption{
Typical reconstruction efficiencies and 
pion misidentification probabilities for 
electrons, muons and kaons in \BaBar. 
} 
\label{tab:pid}
\vspace{0.3cm}
\begin{center}
\begin{tabular}{l|c|c}
\hline
   Particle & Efficiency& $\pi$ contamination  \\
\hline
 electrons & 92\% & 0.3\% \\
 muons     & 75\% & 2.5\% \\
 kaons     & 85\% & 5\% \\
\hline
\end{tabular}
\end{center}
\end{table}
Of particular interest is the good kaon-pion separation capability 
of the detector obtained from the measurement of the Cherenkov angle 
by the DIRC detector. The separation is  better than 3$\sigma$ up to 
momenta of 3.5 GeV/c as illustrated in Figure~\ref{fig:dirc}. 

\begin{figure}
  \begin{center}
  \includegraphics[width=6.7cm]{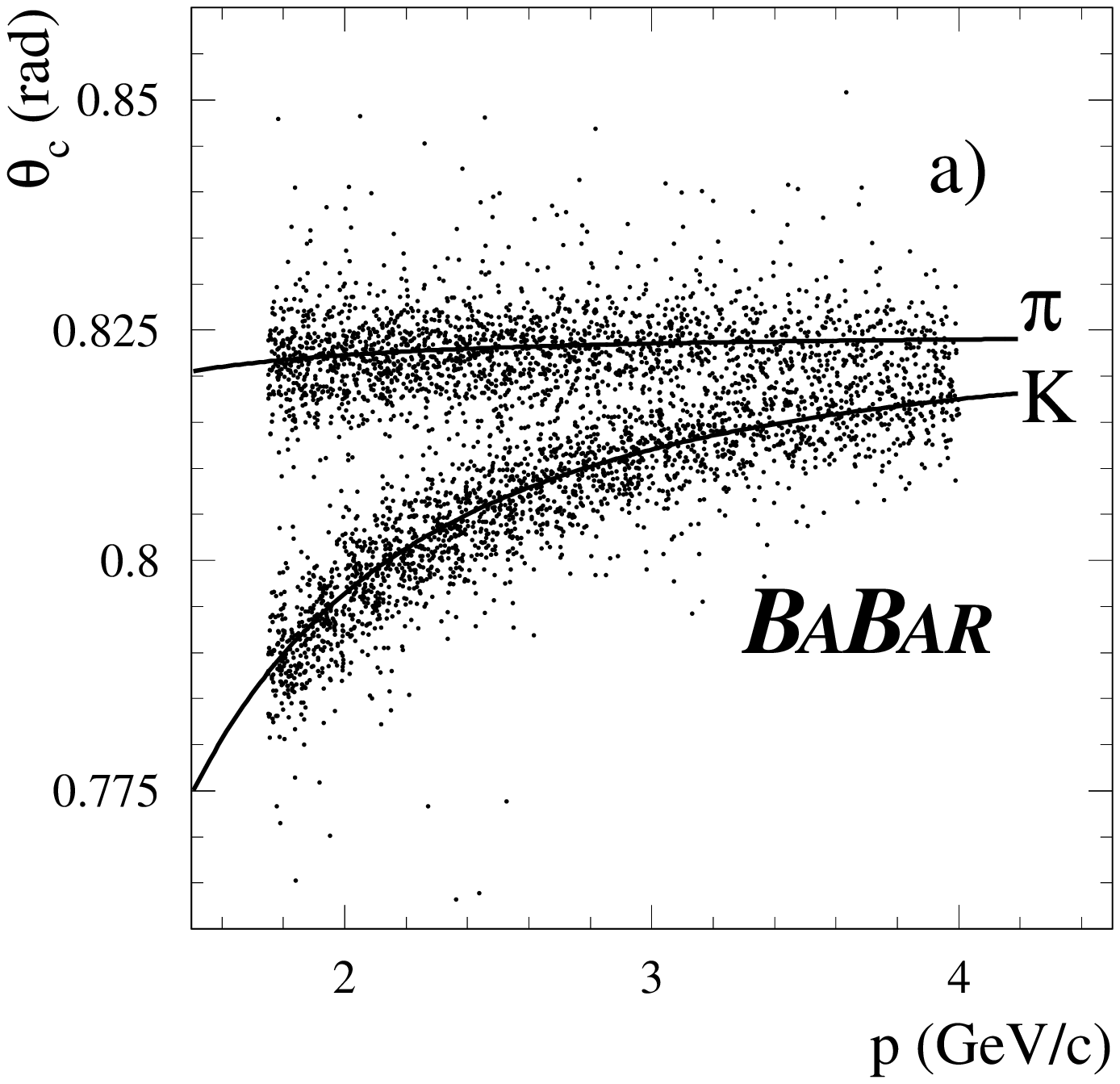}%
  \includegraphics[width=6.7cm]{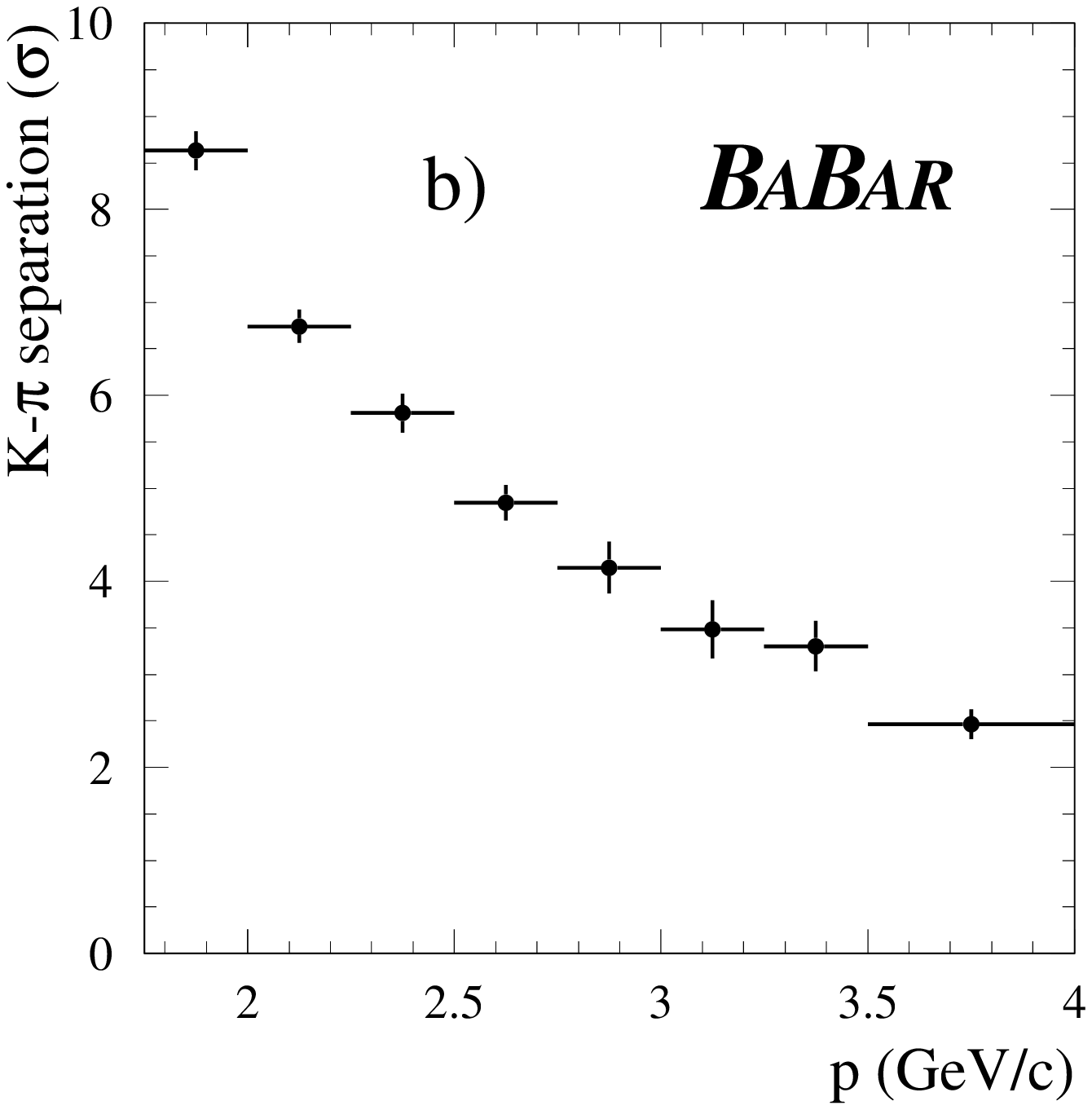}
\vspace{-0.3cm}
  \caption{(a) The Cherenkov angle and (b) $K$--$\pi$ separation as
  functions of momentum for single tracks in a \Dz\ control sample.
  The separation is an average over all polar angles.}
  \label{fig:dirc}
  \end{center}
\end{figure}

A more detailed description of detector design and performance
can be found in~\cite{detector2000}.

\subsection{The data set}

Unless otherwise specified, all the results presented in this talk \
are based on a data set of 
7.7 fb$^{-1}$ recorded at the $\YfourS$ resonance and 
1.2 fb$^{-1}$ recorded off peak. These data were  recorded by the \BaBar\  
detector between  January and June 2000. 

Since the analyses are still in progress, all results are preliminary. 


\section{Study of B $\to$ charmonium decays}

A good understanding of B meson decays into final states
containing a charmonium 
resonance is a prerequisite to an analysis of CP violation in the B system. 
In this section,
an extensive set of measurements of inclusive and 
exclusive branching fractions of B mesons into charmonium  decays 
is briefly summarized. 
More details can be found in~\cite{babar04} 
and~\cite{babar05}. 

\subsection{Inclusive analysis} 


\jpsi\ candidates are formed from pairs of oppositely 
charged tracks both identified as 
leptons. Figure~\ref{fig:jpsimass} shows the invariant  mass distributions 
for $\jpsi\rightarrow\epem$ and $\jpsi\rightarrow\mumu$ candidates after 
 continuum 
subtraction. 
\begin{figure}
  \begin{center}
  \includegraphics[width=13.5cm]{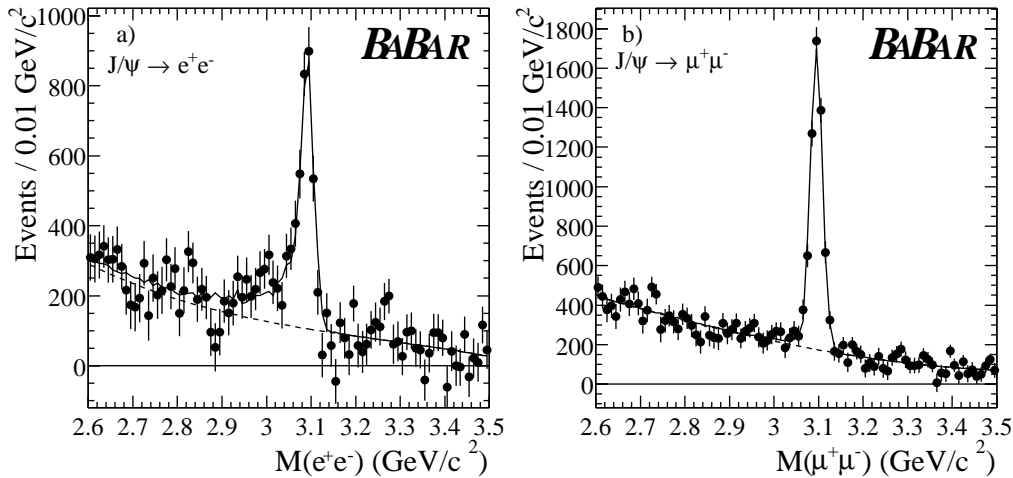}
  \end{center}
\vspace{-0.3cm}
\caption{Mass distribution of \jpsi\ candidates to (a) electron pairs
and (b) muon pairs after the continuum subtraction.}
\label{fig:jpsimass}
\end{figure}
The number of \jpsi\ mesons is extracted from the mass distribution
using a probability density  function (p.d.f.)\ derived from a simulation
that includes final state radiation
and bremsstrahlung. A total of 
$4920 \pm 100 \pm 180$ $\jpsi\rightarrow\epem$ and
$5490 \pm 90 \pm 90$ $\jpsi\rightarrow\mumu$ events were reconstructed. 

The \psitwos\ was reconstructed in its decays to \epem, 
\mumu\ and  \jpsi $\pi^+ \pi^-$. 
For the modes \mumu\ and  \epem, the numbers of signal events are
extracted from the mass distribution  
using a p.d.f.\ from simulation.
A total of  $131 \pm 29$ $  \psitwos\rightarrow\epem$ and
$125 \pm 19$ $\psitwos\rightarrow\mumu$ events were found.
For the decays 
  \jpsi $\pi^+ \pi^-$,
 the yields are 
extracted from a fit to the distribution of
the mass difference between the \psitwos\ and the \jpsi\,  
in order to  reduce the impact of the radiative tail and the mass 
resolution of the \jpsi\ candidate. A total of 
$126 \pm 44$  events were found in candidates with 
$\jpsi\rightarrow\epem$, 
and $162 \pm 23$ events
for $\jpsi\rightarrow\mumu$.

\chic\ mesons are reconstructed through the radiative decay 
$\chic\rightarrow\gamma\jpsi$.  The selected photon candidates 
must have
an energy between 0.20 and 0.65\gev\ in the CM system and be isolated from 
nearby hadronic showers. 
The substantial combinatorial background from $\pi^0$ decays is reduced
by rejecting any photon that, when combined with any other photon,  
gives an  invariant mass compatible 
with a  $\pi^0$.

As for the \psitwos, the numbers of events are extracted by fitting
the distribution of the mass difference between the \chic\ and \jpsi\
candidates (Figure~\ref{fig:chic}). 
We simultaneously fit for a \chicone\ and a possible \chictwo\ component, 
assuming 
the resolution of the two resonances to be the same and fixing
the difference between the \chictwo\ and the \chicone\  masses to the
Particle Data Group value. 
\begin{figure}
\begin {center}
\includegraphics[width=13.5cm]{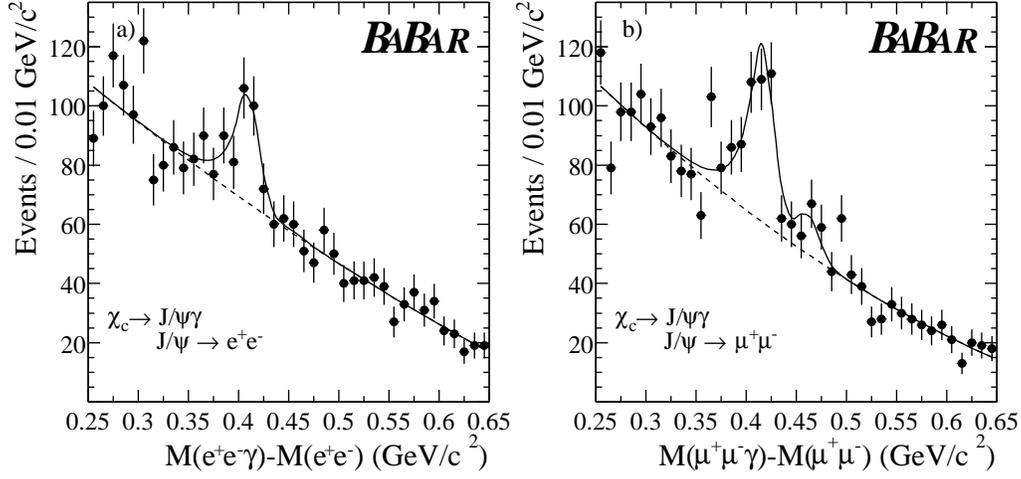}
\end{center}
\caption{Mass difference between the \chic\ and \jpsi\ candidates, 
where $\chic\rightarrow \gamma \jpsi$ and (a) $\jpsi \rightarrow \epem$
and (b) $\jpsi \rightarrow \mumu$.}
\label{fig:chic}
\end{figure}
We found $129 \pm 26 \pm 13$ \chicone\ and $3 \pm 21$ \chictwo\ 
events in which $\jpsi$ decayed into $\epem$, and 
$204 \pm 47 \pm 12$ \chicone\ and $47 \pm 21$ \chictwo\ events
for the $\mumu$ mode.

Due to the preliminary status of the analysis,
we chose not to quote a branching ratio for 
B \to\ \jpsi\ $X$ at this time.
The yields obtained for the decay B \to\ \jpsi\ $X$ are 
used, together with the already precise value of the branching fraction 
for this decay reported in  \cite{pdg2000}, 
as a normalization in the calculation of the branching ratios of  
$B \rightarrow \psi (2s) X$ and $B \rightarrow \chic X$. 
The results,   summarized in Table~\ref{tab:brsum}, are very competitive compared with the world averages. 
\begin{table*}
\caption{Summary of inclusive $B$ branching ratios measured
with respect to $B\rightarrow\jpsi X$.
 The results are combined and multiplied by the 
PDG value for $\BR(B\rightarrow\jpsi X)$ to obtain inclusive
branching fractions (\%).} 
\label{tab:brsum}
\begin{tabular}{l|ccc|ccc|c|ccc} \hline
   & \multicolumn{7}{c|}{Branching ratio relative to \jpsi $X$} &
     \multicolumn{3}{|c}{Branching fraction (\%)} \\ \hline
   & \multicolumn{3}{|c|}{\epem} &  \multicolumn{3}{|c|}{\mumu}
   & Common &  \multicolumn{3}{|c}{Combined} \\ 
   &    & Stat & Sys &      & Stat & Sys & Sys (\%) &     & Stat & Sys \\
   \hline
$\chicone\rightarrow\gamma\jpsi$ & 0.28 & $\pm$ 0.06 &  $\pm$0.03 & 0.38 &  $\pm$0.05 &  $\pm$0.02 &
\phantom{0} $\pm$7.3 & 0.39 &  $\pm$0.04 &  $\pm$0.04 \\ \hline
\chictwo\, 90\% CL & $<$ 0.16 & & & $<$ 0.28 & & & \phantom{0} $\pm$8.1 & $<$ 0.24 & & \\ \hline
$\psitwos\rightarrow\ell^+\ell^-$ & 0.22 & $\pm$0.05 & $\pm$0.01 & 0.23 & $\pm$0.04 & $\pm$0.01 &
$\pm$13.6 & 0.26 & $\pm$0.03 & $\pm$0.04 \\
$\psitwos\rightarrow\pipi\jpsi$ & 0.18 & $\pm$0.06 & $\pm$0.01 & 0.22 & $\pm$0.03 & $\pm$0.00
&\phantom{0}$\pm$9.9 & 0.24 & $\pm$0.03 & $\pm$0.03 \\ 
Combined \psitwos & & & & & & & & 0.25 & $\pm$0.02 & $\pm$0.02 \\ \hline
\end{tabular}
\end{table*}

\subsection{Exclusive analysis}

We considered the exclusive $B$ decay channels
listed in Table~\ref{tab:modes}.
\begin{table}[b]
  \caption[tab:modes]{Exclusive $B$ meson decay modes.}
\vspace{-0.2cm}
   \begin{center}
  \begin{tabular}{l|l}
  \hline
  Channel & Secondary decay mode(s)\\
  \hline
  \Bz$\rightarrow$\jpsi\KS & \jpsi\toLL\\
                           & \KS $\rightarrow$ \pipi, \piz\piz\\
  \BpJKp & \jpsi\toLL\\
  \BNotJKstarz & \jpsi \toLL\\
               & \Kstarz $\rightarrow$ \Kp\pim, \KS\piz\\
  \BpJKstarp & \jpsi \toLL\\
             & \Kstar $\rightarrow$ \KS\pim, \Kp\piz \\
  \BNotpsitwosKs & \psitwos \toLL, \jpsi\pipi\\
                 & \KS $\rightarrow$ \pipi \\
  \BppsitwosKp & \psitwos \toLL, \jpsi\pipi \\
  \BpChiKp  & \chicone  $\rightarrow$ \jpsi $\gamma$; \jpsi \toLL \\
  \hline
  \end{tabular}
  \end{center}
  \label{tab:modes}
\vspace{-0.6cm}
\end{table}

  In these analyses, the reconstruction of the charmonium decays  is very 
  similar to the one used for inclusive analysis,
  but with  
  looser requirements for the lepton identification. 
  In the case of the decay $\jpsi\to\epem$,
  we apply a procedure to add photons 
  which are close to the electron 
  tracks in order to reduce the impact of bremsstrahlung on the reconstruction
  efficiency.
 
  $\KS\rightarrow\pi^+\pi^-$ candidates are 
formed from pairs of oppositely charged tracks 
which have an invariant mass between 0.489 and 0.507\gevcc. 
In addition we 
require that the \KS\ be consistent with
having originated from the \jpsi\ vertex.

$\KS\rightarrow\piz\piz$ candidates are required to have a mass between
0.470 and 0.525\gevcc\ and an energy greater than 0.8\gev. 
A \piz\ decay to two photons is observed in the EMC either as a single 
neutral cluster with substructure
or as two distinct $\gamma$ clusters.
The most probable decay point of the \KS\ is determined after refitting the
two \piz\ mesons at several points along the path defined by their summed 
momentum  vector and the \jpsi\ vertex.  

We reconstruct \Kstarz\ decays to \Kp\pim\ and \KS\piz, and  \Kstarp\ decays to
\KS\pip\ and \Kp\piz. In all cases the candidate $K^{*}$ is required to have an
invariant mass within 0.075\gevcc\ of the nominal value.

Charged kaons are identified by the Cherenkov angle measured by the 
DIRC detector 
and/or by the specific ionization measurement (dE/dx) in the DCH,
depending on particle momentum. 

To isolate the signal for each mode we use the variables \DE, the
difference between the reconstructed and expected $B$ meson energy 
measured in the center-of-mass frame, and \mes, the beam-energy substituted
mass. These variables are defined as:
\begin{eqnarray}
\mes &=& \sqrt{E^{*2}_b - \mbox{\boldmath $p$}_B^{*2}},  \\
\DE &=& E_B^* - E_b^*, 
\end{eqnarray}
where $E_b^*$ is the beam energy in the center-of-mass,
 and $E_B^*$ and
$\mbox{\boldmath $p$}_B^*$ are the energy and momentum of the
reconstructed $B$ meson in the center-of-mass.


\begin{figure*}
  \begin{center}
  \includegraphics[width=12.5cm]{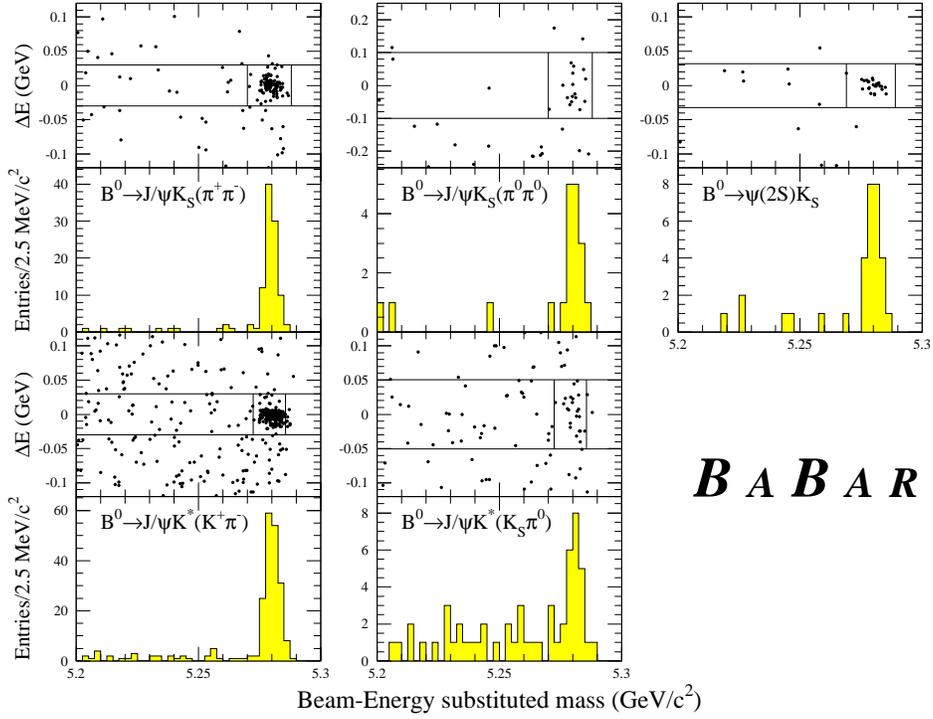}\hfill
  \includegraphics[width=12.5cm]{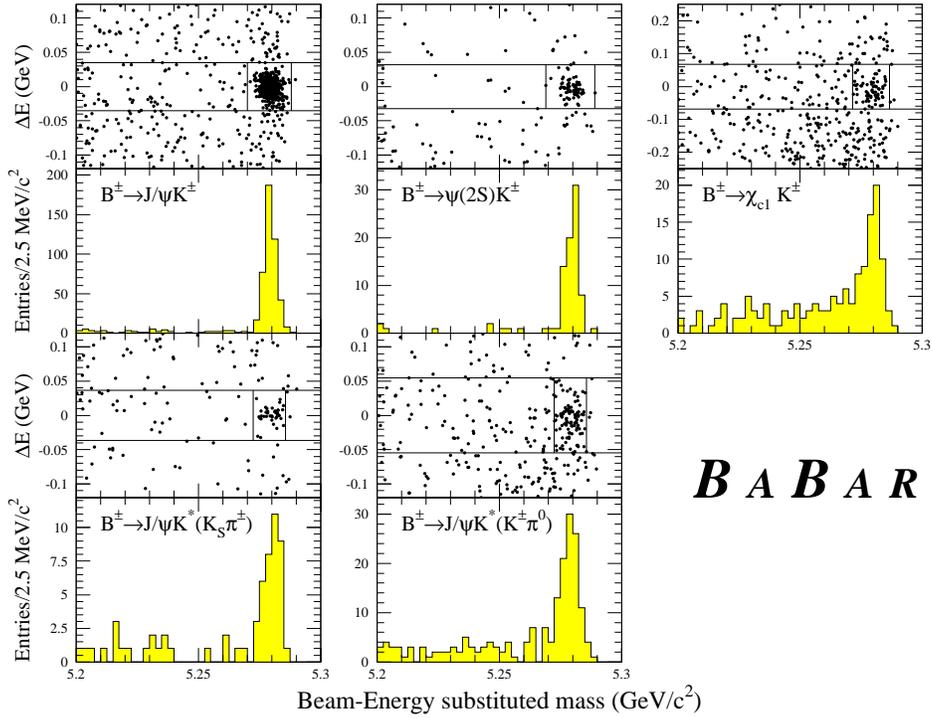}
  \caption{\label{fig:deandmb} 
    Distributions of candidate events in \mes\ and 
    \DE. The upper plot shows the \Bz\ modes and the lower plot the \Bu\
    modes.}
  \end{center}
\end{figure*}

When deriving branching fractions 
we have used the secondary branching fractions and their
associated errors published by the Particle Data Group~\cite{pdg2000}.

We determine the number of \BB\ events from the difference in the
multi-hadron rates  on and off the \FourS\ resonance, normalized
to the respective luminosities.
The efficiencies for each mode have been obtained from Monte Carlo simulations
complemented with measurements of tracking and particle identification
efficiencies extracted from data.
The shape of the beam-energy substituted mass distribution 
is  parameterized for each mode
 with the sum of a Gaussian and the ARGUS function~\cite{ref:ARGUS}.

Figure~\ref{fig:deandmb} shows the \mes\ and \DE\ 
distributions of the candidates.
In Table~\ref{tab:yield} we present the yields 
and measured branching fractions for the individual exclusive
modes.
\begin{table}
  \caption{\label{tab:yield}%
    The yields and measured branching fractions
    for exclusive decays of $B$ mesons involving charmonium.
    The errors on the yields are only statistical.
    For the branching fractions,
    the first error is statistical and the second systematic.}
\vspace {0.4cm}
\begin {center}
  \begin{tabular}{lcc}
    \hline
    Channel & Yield   & Br.~Frac./$10^{-4}$\\
    \hline
    \BNotJKz  \\
             \quad  \KS\ $\rightarrow \pi^+\pi^-$   & 93 $\pm$ 10 & 
             $10.2 \pm 1.1 \pm 1.3 $\\
             \quad  \KS\ $\rightarrow \pi^0\pi^0$ & 14 $\pm$ 4 & 
             $7.5 \pm 2.0 \pm 1.2  $\\
    \BpJKp &             445 $\pm$ 21 & 
             $11.2 \pm 0.5 \pm 1.1  $\\
    \BNotJKstarz &       188  $\pm$ 14 & 
             $13.8 \pm 1.1 \pm 1.8  $ \\
    \BpJKstarp &   126 $\pm$ 12  &
             $13.2 \pm 1.4 \pm 2.1  $\\ 
    \BNotpsitwosKz &  23 $\pm$ 5 & 
             $8.8 \pm 1.9 \pm 1.8  $\\
    \BppsitwosKp & 73  $\pm$ 8 & 
             $6.3 \pm 0.7 \pm 1.2  $  \\
    \BpChiKp &  44 $\pm$ 9 & 
             $7.7 \pm 1.6 \pm 0.9  $ \\
    \hline
  \end{tabular}
\end{center}
\vspace {-0.1cm}
\end{table}
Figure~\ref{fig:brsum} shows the measured branching fractions compared
to the values compiled by the Particle Data Group~\cite{pdg2000}.
\begin{figure}
  \begin{center}
  \includegraphics[width=10.5cm]{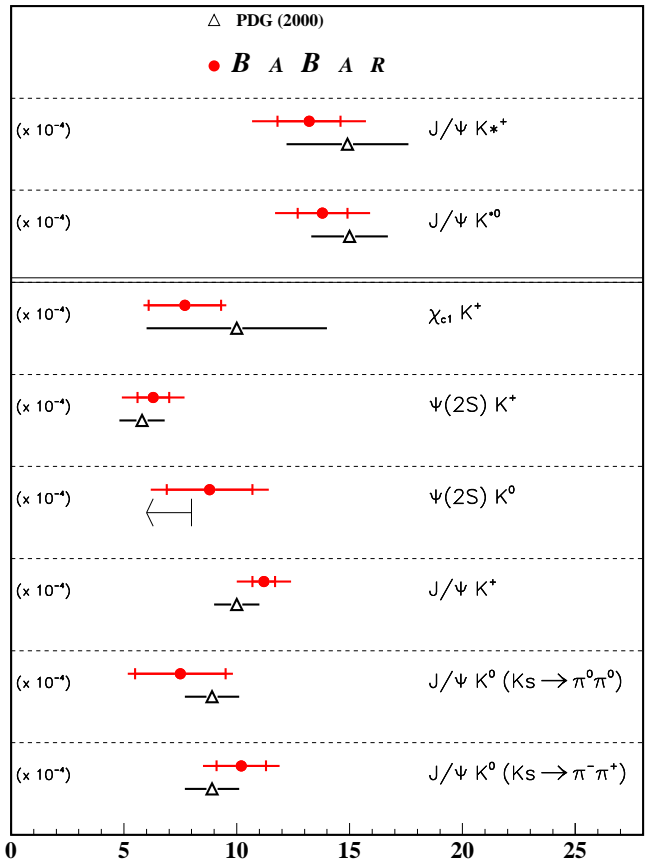}
\vspace{-0.3cm}
  \caption{\label{fig:brsum} 
  Summary of branching fraction measurements for charmonium + $K$ channels 
  and comparisons with the PDG 2000 values.}
  \end{center}
\vspace {-0.5cm}
\end{figure} 


\section{Measurement of $\sin2\beta$}

The measurement of \stwob\ requires five basic steps: 
exclusive reconstruction of the CP mode; 
measurement of time elapsed between the decays of the two B mesons;   
tagging of the flavour of the B meson at production time; 
measurement of the dilution factors from the data;
extraction of the value of \stwob\ from a fit to the 
       amplitude of the \CP\ asymmetry.    
The following subsections briefly describe the analysis strategy and
the result. 
For a more detailed description see \cite{babar01} and references  
therein.

\subsection{Reconstruciton of the CP mode}
For a first measurement,
we targeted only decays characterized by relatively high 
accessible branching fractions ($\sim$10$^{-4}$--10$^{-5}$),
low backgrounds ($<$10\%),
and reliable theoretical prediction (i.e., no penguin pollution). 
Only the decays $B \to J/\psi K_s$ and $B \to \psi(2S) K_s$
have been used so far. 
The procedure used for the exclusive reconstruction of the B meson is
described in~\cite{babar05}.  

The yields and purities 
obtained in each channel analysing 9 fb$^{-1}$
are reported in Table~\ref{tab1}. 
\begin{table}[b]
\caption{The CP sample used for the \stwob\ measurement:
yields and purities 
obtained using 9 fb$^{-1}$.}
\label{tab1}
\begin {center}
\begin{tabular}{lll}
\hline
Decay mode                & Yield            & Purity \\
\hline
$J/\psi K_s (K_s \to \pi^+ \pi^-)$  & 124 $\pm$ 12 & 96\% \\
$J/\psi K_s (K_s \to \pi^0 \pi^0) $ & 18 $\pm$ 4  & 91\% \\
$\psi(2s) K_s (K_s \to \pi^+ \pi^-)$  & 27 $\pm$ 6  & 93\% \\
\hline
\end{tabular}
\end{center}
\end{table}
The modes $B \to \chi_c K_s$, $B \to J/\psi K_L$
and $B \to J/\psi K^{*0}$ are still under  
investigation and will be included in the analysis in the future. 

\subsection{Time resolution function }
The resolution of the \deltat\ measurement 
is dominated by the $z$ resolution of the tagging vertex.
The three-momentum of the tagging $B$ and its associated 
error matrix are derived from the fully reconstructed 
 $B_{CP}$ candidate three momentum, 
decay vertex and
error matrix, and from the knowledge of the average position
of the interaction point and the \FourS\ four-momentum.
This derived $B_{tag}$ three-momentum is fit to a common vertex with the 
remaining tracks in the event (excluding those from $B_{CP}$).
In order to reduce 
the bias due to long-lived particles, all reconstructed $V^0$ 
candidates are used as input to the fit in place
of their daughters.  Any track whose contribution to the $\chi^2$
is greater than 6 is removed from the fit. 
This procedure is iterated until there
are no tracks contributing more than 6 to the $\chi^2$ 
or until all tracks are removed.   
\par
The time resolution function is described  by the sum of
 two Gaussian distributions:
\begin{equation}
{\cal {R}}(\deltat ) =
\sum_{i=1}^{2} {  \frac{f_i}{\sigma_i\sqrt{2\pi}} \exp  \left(
- \frac{( \deltat-\delta_i )^2}{2{\sigma_i }^2}   \right) }.
\end{equation}
Monte Carlo simulation 
indicates that 
$70\%$ of the events are in the core 
Gaussian, which has a width $\sigma_1 \approx 0.6 \ps$.
The wide Gaussian has a width  $\sigma_2 \approx 1.8\ps$. Tracks from forward-going charm decays included in the reconstruction of the $B_{tag}$ vertex 
introduce a small bias, 
$\delta_1 \approx -0.2 \ps$, for the core Gaussian.  
\par
A small fraction of events ($\sim$1\%)  
have very large values of \deltaz, 
mostly due to vertex reconstruction
problems.  This is  accounted for in the 
parameterization of the time resolution function
with a very wide unbiased Gaussian with fixed width of $8\ps$.   
\par 
In likelihood fits, we use a parameterization of the 
error $\sigma_{\deltat}$ 
calculated from the fits to the two \B\ vertices for each individual event.
 We introduce two scale factors ${\cal S}_1$ and ${\cal S}_2$ 
for the width of the narrow and the wide Gaussian distributions
($\sigma_1={\cal S}_1 \times \sigma_{\deltat}$ and 
$\sigma_2={\cal S}_2 \times \sigma_{\deltat}$) to account for the fact
that the uncertainty on \deltat\ is underestimated due to effects such as
the inclusion of particles from $D$ decays and possible underestimation 
of the amount of material traversed by the particles.
The scale factor ${\cal S}_1$ and the bias $\delta_1$ of the 
narrow Gaussian are free parameters in the fit.  
The scale factor ${\cal S}_2$ and the fraction of events in the wide
Gaussian, $f_2$, are fixed to the values estimated 
from Monte Carlo  simulation by a fit to the pull distribution (${\cal S}_2=2.1$
and $f_2=0.25$). The bias of the wide Gaussian, $\delta_2$, 
is fixed at $0 \ps$.
The last free parameter in the fit is
$f_w$, the fraction of the very wide Gaussian that describes
the tail of the resolution.

Because the time resolution is dominated by the precision 
of the $B_{tag}$ vertex position, we find  no significant 
differences in the Monte Carlo simulation of 
the resolution function parameters 
for the various fully reconstructed decay modes.
This allows us to determine
the resolution function parameters
with the relatively high-statistics fully-reconstructed \Bz\ data samples.  
The differences in the resolution function parameters in the 
different tagging categories (see below) are also small.  

Table~\ref{tab:Resolution} presents the values of the \deltat\ resolution
parameters obtained 
from a maximum likelihood fit to a sample of fully reconstructed \Bz\ decays. 
Further
details on the procedure and the results
can be found in~\cite{babar08}.
The vertex parameters are fixed to these values in
the final unbinned maximum likelihood 
fit for \stwob\ in the low-statistics \CP\ event sample. 
\begin{table}
\caption{
Parameters of the resolution function determined from the sample
of events with fully-reconstructed hadronic \Bz\ candidates.
} 
\label{tab:Resolution}
\vspace{0.2cm}
\begin {center}
\begin{tabular}{cc|cl}
\hline
   \multicolumn{2}{c|}{Parameter} & \multicolumn{2}{c} {Value} \\
\hline
 $\delta_1$  & (ps)    & $-0.20\pm0.06$  & from fit     \\
 ${\cal S}_1$&   & $1.33\pm0.14$       & from fit     \\
 $f_{w}$       & (\%)  & $1.6\pm0.6$     & from fit     \\
 $f_1$       & (\%)  & $75$              & fixed        \\
 $\delta_2$  & (ps)  & $0$             & fixed        \\
 ${\cal S}_2$ &  & $2.1$               & fixed        \\
\hline
\end{tabular}
\end{center}
\end{table}

\subsection{B flavor tagging} 
\label{sec:Tagging}

Each event with a \CP\ candidate is assigned a $\Bz$ or $\Bzb$ tag if 
the rest of the event (i.e., with the daughter
tracks of the $B_{CP}$ removed) satisfies the criteria for one of several 
tagging categories. 
The figure of merit for each tagging category is the effective tagging
efficiency  
$Q_i = \varepsilon_i \, \left( 1 - 2\mistag_i \right)^2$, where $\varepsilon_i$ 
is the fraction of events assigned to category $i$ and 
$\mistag_i$ is the probability of misclassifying the tag as a $\Bz$ or $\Bzb$
for that category. $\mistag_i$ are called  mistag fractions.
The statistical error
on \stwob\ is proportional to $1/\sqrt{Q}$, where $Q = \sum_i Q_i$. 

The algorithm used in the analysis categorizes the events in four different 
classes: 
\begin{description} 
\item[{\tt Lepton}] when the flavor of the B meson is identified by the charge of 
      the high momentum lepton coming from a semileptonic decay;    
\item[{\tt Kaon}] when the flavor of the B meson is identified by the charge of 
      the kaon coming from the hadronization of the s quark coming from the decay $b\to c\to s$; 
\item[{\tt NT1}]  the tag is chosen according to the output of a neural network that exploits  
            other (correlated) information, such as secondary lepton charge, slow pions from D$^*$ decays, 
            jet charge;     
\item[{\tt NT2}]  similar to NT1 but characterized by a smaller tagging power. 
\end{description}
When an event satisfied the selection criteria for more than one tagging category, it was 
assigned to the tag with the highest discrimination power. 

\subsubsection {Measurement of mistag fractions } 
The mistag fractions are measured directly
in events 
in which one \Bz\ candidate, called the $B_{rec}$, is fully reconstructed 
in a flavor eigenstate mode.
The flavor-tagging algorithms described in the previous section 
are applied to the 
rest of the event, which constitutes the potential $B_{tag}$.
\par
Considering the \BzBzb\ system as a whole, one can classify the tagged events 
as {\em mixed} or {\em unmixed} depending on whether the $B_{tag}$ is tagged 
with the same flavor as the $B_{rec}$ or  with the opposite flavor.   
The observed fraction of mixed events at the time $t$ is expressed as 
\begin{equation}
\label{eq:TagMix:Timedependent}
\chi(\Delta t) = \mistag + D/2 ( 1-\cos (\Delta m \Delta t)), 
\end{equation}
were $D = 1 - 2w$ is the so called $dilution factor$.  
The mixing probability is smallest for small values of
$\deltat = t_{rec}-t_{tag}$ 
so that the apparent rate of mixed events near $\deltat=0 $
is governed by the mistag probability.
In addition to improving sensitivity to the mistag fraction,
this time-dependent measurement technique
also helps discriminate against backgrounds with 
different time-dependence.

The extraction of the mistag probabilities
is complicated in reality by the possible presence of mode-dependent 
backgrounds. 
We deal with these by adding specific terms in the likelihood 
functions describing the different types of backgrounds (zero lifetime, 
non-zero lifetime without mixing, non-zero lifetime with mixing).
Details are described in~\cite{mhs} and~\cite{babar08}.  

\begin{table*}
\caption{
Categories of tagged events in the \CP\ sample.
The {\tt Lepton} category is split into
{\tt Electron} and {\tt Muon} categories.
} 
\begin{tabular}{l|ccc|ccc|ccc|ccc} \hline
  & \multicolumn{6}{c|}{\rule[-1pt]{0mm}{14pt}$\jpsi \KS$} 
  & \multicolumn{3}{c|}{$\psitwos \KS$} 
  & \multicolumn{3}{c}{\CP\ sample } \\ \cline{2-13}
Tagging Category 
  &  \multicolumn{3}{c|} {\rule[-1pt]{0mm}{14pt}($\KS \to \pi^+\pi^-$)} 
  &  \multicolumn{3}{c|} {($\KS \to \pi^0\pi^0$)} 
  &  \multicolumn{3}{c|} {($\KS \to \pi^+\pi^-$)} 
  &  \multicolumn{3}{c}  {(tagged)} 
  \\ \cline{2-13}
     &  \Bz\ & \rule[-1pt]{0mm}{14pt}\Bzb\ & all &  \Bz\ & \Bzb\ & all &  \Bz\ & \Bzb\ & all &  \Bz\ &
 \Bzb\ & all \\
 \hline
{\tt Electron}   
  & 1  &  3 & 4  
  & 1  &  0 & 1  
  & 1  &  2 & 3 
  & 3  &  5 & 8      \\
{\tt Muon}       
  & 1  &  3 & 4  
  & 0  &  0 & 0  
  & 2  &  0 & 2 
  & 3  &  3 & 6      \\
{\tt Kaon}       
  & 29 & 18 & 47 
  & 2  & 2  & 4  
  & 5  & 7  &12 
  & 36 & 27 & 63     \\
{\tt NT1}       
  &  9 &  2 & 11 
  & 1  & 0  & 1  
  & 2  & 0  & 2 
  & 12 & 2  & 14     \\
{\tt NT2}       
  & 10 &  9 & 19 
  & 3  & 3  & 6  
  & 3  & 1  & 4 
  & 16 & 13 &  29    \\
\hline
{\tt Total}      
  &  50& 35 & 85 
  & 7  & 5  &12 
  &13  &10  &23  
  & 70 & 50 & {\bf 120 }   \\
\hline
\end{tabular}
\label{tab:TaggedEvents}
\end{table*}

\begin{table}
\caption{
Mistag fractions measured from 
a maximum-likelihood fit to the time distribution for the fully-reconstructed \Bz\ sample.
The uncertainties on $\varepsilon$ and $Q$ are statistical only.
} 
\label{tab:TagMix:mistag}
\vspace{0.2cm}
\begin{center}
\begin{tabular}{l|c|c|c}
\hline
Category & $\varepsilon$ (\%) & $\mistag$ (\%) & $Q$ (\%)     \\
\hline
{\tt Lepton}     & $11.2\pm0.5$ & $9.6\pm1.7\pm1.3$   &  $7.3\pm0.7$  \\
{\tt Kaon}       & $36.7\pm0.9$ & $19.7\pm1.3\pm1.1$  &  $13.5\pm1.2$  \\
{\tt NT1}        & $11.7\pm0.5$ & $16.7\pm2.2\pm2.0$  &  $5.2\pm0.7$  \\
{\tt NT2}        & $16.6\pm0.6$ & $33.1\pm2.1\pm2.1$  &  $1.9\pm0.5$  \\
\hline
all              & $76.7\pm0.5$ &                     &  $27.9\pm1.6$ \\ 
\hline
\end{tabular}
  \end{center}
\vspace{-0.3cm}
\end{table}

The mistag fractions and the tagging efficiencies 
obtained by combining the results from maximum likelihood fits to the
time distributions in the \Bz\
hadronic and semileptonic samples are summarized 
in Table~\ref{tab:TagMix:mistag}. We find a tagging efficiency of 
$(76.7\pm0.5)\%$ (statistical error only).
The lepton categories have the lowest mistag fractions, 
but also have a low efficiency.  
The {\tt Kaon} 
category, despite having a larger mistag fraction (19.7\%),
has a higher effective tagging efficiency;
one-third of events are assigned to this category.   
Altogether, lepton and kaon categories have an effective tagging 
efficiency $Q \sim 20.8\%$.  
The neural network categories increase the 
effective tagging efficiency by $\sim 7\%$ to an overall 
$Q = (27.9 \pm 0.5) \%$ (statistical error only).

Of the 168 \CP\ candidates, 120 are tagged:  70 as \Bz\ and 50 as \Bzb.
The number of tagged events per category is given in Table~\ref{tab:TaggedEvents}.

\subsection{Results}

The maximum-likelihood fit for \stwob,
using the full tagged sample of $\Bz/\Bzb \to \jpsi \KS$
and $\Bz/\Bzb \to \psitwos \KS$ 
events, gives:
\begin{equation}
\stwob=0.12\pm 0.37 {\rm (stat)} \pm 0.09 {\rm (syst)}\, .
\end{equation}
For this result,
the \Bz\ lifetime and \deltamd\ are fixed to the current best
values~\cite{pdg2000},
and \deltat\ resolution parameters and the mistag rates are fixed to the values obtained from data as summarized
in Tables~\ref{tab:Resolution} and \ref{tab:TagMix:mistag}.
\begin{figure}
\begin{center}
\includegraphics[width=11cm]{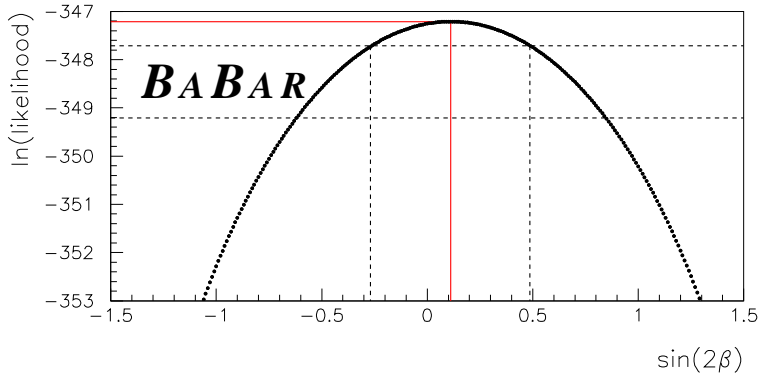}
\end{center}
\vspace{-0.4cm}
\caption{ 
Variation of the log likelihood as a function of \stwob. 
The two horizontal dashed lines indicate changes in the log likelihood 
corresponding to one and two statistical standard deviations.
}
\label{fig:likelihood}
\end{figure}
The log likelihood is shown as a function of \stwob\ 
in Figure~\ref{fig:likelihood}. 

\begin{table}
\caption{
Result of fitting for \CP\ asymmetries in the entire \CP\ sample and in 
various subsamples.
} 
\vspace{0.3cm}
\begin{center}
\begin{tabular}{l|c} \hline
 sample                                    &  \stwob  \\ \hline
 \CP\ sample                               &  {\bf 0.12}$\pm${\bf 0.37}  \\  
\hline
 \ \ $\jpsi \KS$ ($\KS \to \pi^+ \pi^-$) events  &  $-0.10 \pm 0.42$   \\  
 \ \ other \CP\ events                           &  $0.87 \pm 0.81$   \\  
\hline
 \ \ {\tt Lepton}                         &  $1.6 \pm 1.0  $  \\
 \ \ {\tt Kaon}                           &  $0.14\pm 0.47   $   \\
 \ \ {\tt NT1}                            &  $-0.59\pm0.87  $   \\
 \ \ {\tt NT2}                            &  $-0.96\pm 1.30  $   \\
\hline 
\end{tabular}
\end{center}
\label{tab:result}
\end{table}
The results of the fit for each
type of \CP\ sample and for each tagging category are given in Table~\ref{tab:result}.  
\begin{table}
\caption{
Summary of systematic uncertainties.  
The different contributions to the systematic 
error are added in quadrature.
} 
\vspace{0.4cm}
\begin {center}
\begin{tabular}{l|c} \hline
                          
 Source of uncertainty    & Error on  \stwob \\ \hline
 
 $\tau_\Bz$                         &   0.002     \\
 \deltamd                           &   0.015     \\
 \deltaz\ resolution                &   0.019     \\ 
 time-resolution bias               &   0.047     \\ 
 measurement of mistag fractions        &  0.053    \\   
 different mistag fractions & \\
 \quad for \CP\ and non-\CP\ samples       &   0.050     \\
 \quad for \Bz\ and \Bzb\     &   0.005     \\
 background                         &   0.015     \\
\hline
 total systematic error                            & {\bf 0.091 }   \\ 
\hline 
\end{tabular}
\label{tab:systematics}
\end {center}
\end{table}
The contributions to the systematic uncertainty are summarized
in Table~\ref{tab:systematics}. 

We estimate the probability of obtaining
the observed value of the statistical uncertainty, 0.37, 
on our measurement of \stwob\
by generating a large number of toy Monte Carlo experiments
with the same number of tagged \CP\ events, and distributed in the same 
tagging categories,  as in the \CP\ sample in the data.  
We find that the errors are distributed
around $0.32$ with a standard deviation  of $0.03$,
and that the probability of obtaining 
a value of the statistical error larger than the one we observe is 5\%.
Based on  a large number of full Monte Carlo simulated experiments with the 
same number of events as our data sample, we 
estimate that the probability of finding a lower value of the likelihood than
our observed value is 20\%.

To validate the analysis we use the charmonium 
control sample,  composed of $B^+ \to \jpsi K^+$ events
and events with self-tagged  $\jpsi K^{*0}$  ($K^{*0} \to K^+ \pi^-$)
neutral \B's.  We also use the event samples with fully-reconstructed
candidates in charged or neutral hadronic modes.
These samples should exhibit no time-dependent asymmetry. 
In order to 
investigate  this experimentally, we define an ``apparent 
\CP\ asymmetry'',  analogous to \stwob\  in 
Eq.~\ref{eq:asym}, which we extract from the data using an identical
maximum-likelihood procedure. 
\par 
The events in the control samples are flavor eigenstates and not
\CP\ eigenstates.
They are used  for testing 
the fitting procedure with the same tagging algorithm as for the \CP\
sample and, in the case of the $B^+$ modes, with self-tagging based on 
their charge.
We also perform the fits for \Bz\ and \Bzb\ (or $B^+$ and $B^-$) events
separately to study possible flavor-dependent systematic effects.    
For the charged \B\ modes, we use mistag fractions measured
from the sample of hadronic charged \B\ decays.
\par
In all fits, including the fits to charged samples, 
we fix the lifetime  $\tau_{\Bz}$ and the oscillation frequency  \deltamd\ 
to the PDG values~\cite{pdg2000}. 
The results of a series of validation checks on the control samples
are summarized in Table~\ref{tab:validation}.
\begin{table}
\caption{
Results of fitting for apparent \CP\ asymmetries in various 
charged or neutral flavor-eigenstate \B\ samples. 
} 
\vspace{0.3cm}
\begin{center}
\begin{tabular}{l|c}
\hline
 Sample                                    &  Apparent \CP-asymmetry  \\ 
\hline
 Hadronic charged \B\ decays                  &   $0.03 \pm 0.07$   \\  
 Hadronic neutral \B\ decays            &   $-0.01 \pm 0.08$  \\  
 $\jpsi K^+$                               &   $0.13\pm 0.14$    \\ 
 $\jpsi K^{*0}$ ($K^{*0} \to K^+ \pi^-$)   &  $ 0.49 \pm 0.26$   \\ 
\hline 
\end{tabular}
  \end{center}
\label{tab:validation}
\vspace{-0.3cm}
\end{table}
The two high-statistics samples and the $\jpsi K^+$ sample 
give an apparent \CP\ asymmetry consistent with zero.  
The 1.9$\sigma$ 
asymmetry in the $\jpsi K^{*0}$ 
is interpreted
as a statistical fluctuation. 

Other \babar\ time-dependent analyses presented at this Conference 
demonstrate the validity of the novel technique developed for 
use at an asymmetric \BF.
In particular, the measurement of the \Bz-\Bzb\ oscillation 
frequency described in~\cite{babar08} uses the same 
time resolution function and tagging algorithm as the \CP\ analysis, 
and the \Bz\ lifetime measurement described in~\cite{babar07}
uses the same inclusive 
vertex reconstruction technique as the \CP\ analysis.
Both result are consistent with the world average~\cite{pdg2000}. 

\subsection{Current knowledge of the angle $\beta$}

Figure~\ref{fig:UnitarityTriangle} shows 
the ($\bar{\rho},\bar{\eta})$ plane,
with \babar's measured central value of 
\stwob\ shown as two straight lines.
There is a two-fold ambiguity in deriving a value of $\beta$
from a measurement of \stwob.
Both choices are shown with cross-hatched regions corresponding
to $1\sigma$ and $2\sigma$ experimental uncertainty.
The ellipses correspond to the regions allowed by all other
measurements that constrain the Unitarity Triangle for a
variety of representative theoretical parameters.
This procedure is discussed in detail in~\cite{MSConstraints}.
The following set of measurements and model-dependent parameters are used:
$\left| V_{cb} \right| = 0.0402\pm0.017$;
$ \left| V_{ub}/V_{cb} \right| = \left< \left| V_{ub}/V_{cb} \right| \right> \pm 0.0079$;
$\deltamd = 0.472\pm0.017 \, \hbar \ps^{-1}$;
$\left| \epsilon_K  \right| = \left( 2.271 \pm 0.017 \right) \times 10^{-3}$;
the set of amplitudes corresponding to a $95\%$CL limit of $14.6 \, \hbar \ps^{-1}$ for ${\rm \Delta} m_s$;
$\left< \left| V_{ub}/V_{cb} \right| \right>$ in $\left[ \, 0.070, \, 0.100 \, \right]$;
$B_K$ in $\left[ \, 0.720, \, 0.980 \, \right]$;
$f_{B_d} \sqrt{ B_{B_d} } $ in $\left[ \, 185, \, 255 \, \right] \mev$;
and $\xi_s$ in $\left[ \, 1.07, \, 1.21 \, \right]$.

\begin{figure}
\begin{center}
\includegraphics[width=10.5cm]{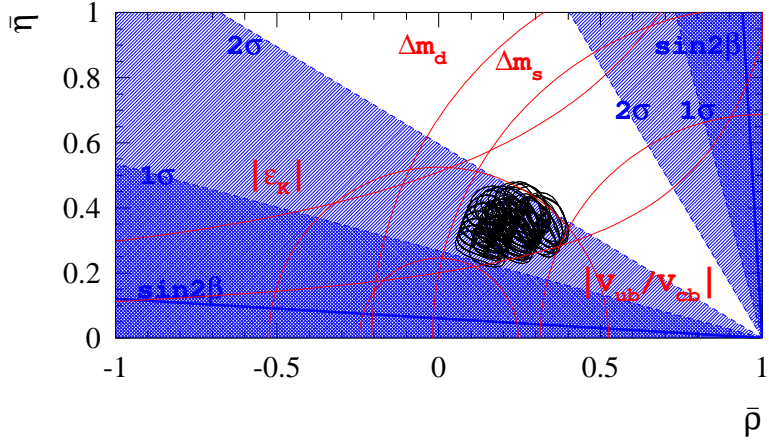}
\vspace{-0.6cm}
\caption{ 
Present constraints on the position of the apex of the Unitarity Triangle in the ($\bar{\rho},\bar{\eta})$ plane.
Our result $\stwob = 0.12\pm0.37{\rm (stat)}$ is represented by cross-hatched regions corresponding to 
one and two statistical standard deviations.
The fitting procedure is described in Ref.~\cite{MSConstraints}.   
}
\label{fig:UnitarityTriangle}
\end{center}
\end{figure}
\begin{figure}
\begin{center}
\includegraphics[width=10.0cm]{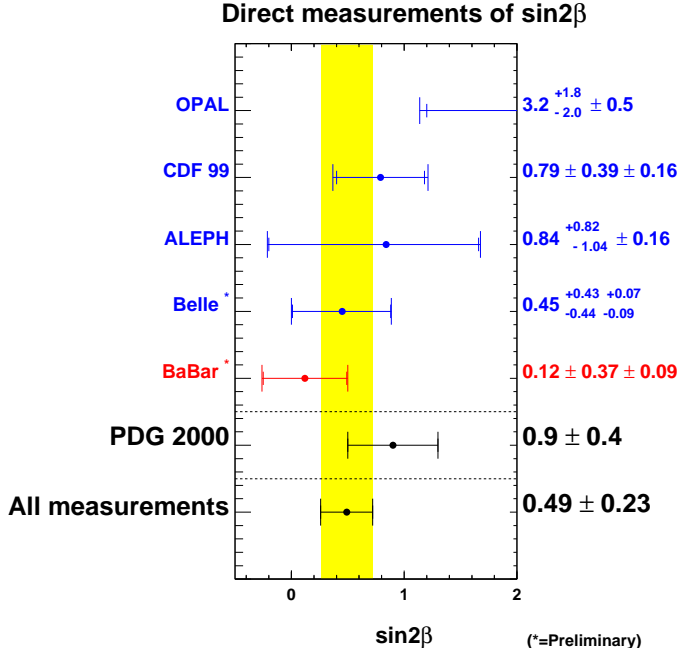}
\caption{
Existing direct measurements of $\sin2\beta$. 
The input measurements are in Refs.~\cite{opal}, 
\cite{cdf99}, 
\cite{aleph}, 
\cite{belle}, 
\cite{babar01}. 
}
\label{fig:average}
\end{center}
\end{figure}
In Figure~\ref{fig:average},
a summary of all existing direct measurements of $\sin2\beta$ is 
reported. The PDG 2000 world average,
obtained from the OPAL~\cite{opal} and CDF~\cite{cdf99} results, 
 is compared with a new average that includes the 
recent ALEPH~\cite{aleph}, Belle~\cite{belle} and \BaBar~\cite{babar01} 
results.
The new measurements 
from the two B factories improved the knowledge of $\sin2\beta$
by a factor 2.

\section{Study of B $\to$ charmless 2-body decays}

\subsection{Interest of the measurement} 
The branching fractions for the rare charmless decays 
$\Bz\to h^+h^-$ $(h=\pi,K)$
provide important information in the 
study of {\CP} violation.
If the tree level diagram dominates the decay, 
the {\pip\pim} decay 
mode can be used to extract the angle $\alpha$ of the Unitarity Triangle. 
If the penguin contribution is not negligible, 
large theoretical uncertainties~\cite{ref:gronau1}
complicate the analysis and the 
extraction of $\alpha$ will require a full isospin 
analysis~\cite{ref:gronau2}.
On the other hand, large penguin amplitude improves the 
prospects for observing direct {\CP} violation as an asymmetry in the 
decay rates for $\Bz\to\Kp\pim$ and $\Bzb\to\Km\pip$.  

The decay $\Bz\to\Kp\pim$ is dominated by the $b\to sg$ penguin amplitude,
and provides an estimate of the scale of penguin pollution in the $\pip\pim$
decay.
Precise measurement of the decay rates for \pipi\ and \kpi\ decays is 
therefore of central importance.

\subsection{Analysis strategy } 
The first step of this analysis consists in
selecting decays B \to $h^+h^-$, where 
$h^{\pm}$ is either a pion or a kaon. 
\B\ decays to charm mesons are not
a significant background to charmless two-body decays
because of the relatively small CM momenta of decay products
produced from the $b\to c$ transition.
The background is dominated by the continuum production of light quarks, 
$\epem\to\qqbar$ $(q = u, d, s, c)$,
which typically exhibits a two-jet structure in the CM frame.
The topology is markedly different from that of real
$B\to h^+h^-$ events, which is more spherically symmetric.
This difference is exploited using the angle $\theta_S$
between the sphericity axes, evaluated in the CM frame, of the \B\ candidate 
and the remaining charged and neutral particles in the event. 

Further separation between signal and continuum background is provided by a
Fisher discriminant technique~\cite{cleo2}.  The Fisher discriminant \fish\ is 
calculated from a linear combination of 9 discriminating variables
constructed from the scalar sum of the momenta of all charged 
and neutral particles (excluding the candidate daughter tracks) flowing 
into 9 concentric cones centered on the $B$-candidate thrust axis in the CM
frame. 
 More energy will be 
found in the cones nearer the candidate thrust axis in jet-like continuum 
background events than in the more isotropic \BB\ events.


Once the decay $B \to h^+h^-$ is identified, the separation between different 
channels is obtained  using two different approaches: 
\begin {itemize}
\item likelihood analysis, in which the $K$-$\pi$ separation is determined
on a statistical basis; 
\item cut based analysis, in which the $K$-$\pi$ separation is performed
track-by-track.
\end {itemize} 
A good understanding of the particle identification performances 
is essential in both analyses.

\subsection{Results} 

We determine branching fractions for $\pip\pim$ and $\Kp\pim$ decays and 
an upper limit for the $\Kp\Km$ decay using the results of the global 
likelihood fit.  
The results are summarized in Table~\ref{tab:brresult}.
\begin{table}
\caption{%
Summary of $B\to\mbox{two-body}$ branching fraction results.
Shown are the central fit values $N_S$, the statistical significance,
and the measured branching fractions $\BR$.
For the $\kk$ mode, the $90\%$ confidence level upper limits are given.
The first errors are statistical and the second systematic.}
\vspace{0.4cm}

\begin{center}
\begin{tabular}{cc@{\ }c@{\ }c} \hline
 Mode & $N_S$ & Stat. Sig. & $\BR\,(10^{-6})$ \\\hline
 \smallskip
 $\pip\pim$ & $29 ^{+8}_{-7}$$^{+3}_{-4}$ & 5.7$\sigma$ & $9.3^{+2.6}_{-2.3}$$^{+1.2}_{-1.4}$ \\
 \smallskip
 $\Kp\pim$  & $38 ^{+9}_{-8}$$^{+3}_{-5}$ & 6.7$\sigma$ & $12.5^{+3.0}_{-2.6}$$^{+1.3}_{-1.7}$ \\
 \smallskip
 $\Kp\Km$   & $7 ^{+5}_{-4}$ ($<15$) & 2.1$\sigma$ & $<6.6$ \\
 \hline
\end{tabular}
\end{center}
\label{tab:brresult}
\end{table}
For the $\kk$ mode we calculate the $90\%$ confidence level upper limit yield 
and decrease the efficiency by the total systematic error $(24\%)$ before
calculating the upper limit branching fraction.
The statistical significance of a given signal yield is determined by setting the yield 
to zero and maximizing the likelihood with respect to the remaining variables. 

Figure~\ref{fig:sig} shows the $n\sigma$ likelihood contour 
curves, where $\sigma$ represents the statistical uncertainty only.  The curves are 
computed by maximizing the likelihood with respect to the remaining variables in the fit.

\begin{figure}
\begin{center}
\includegraphics[width=8.0cm]{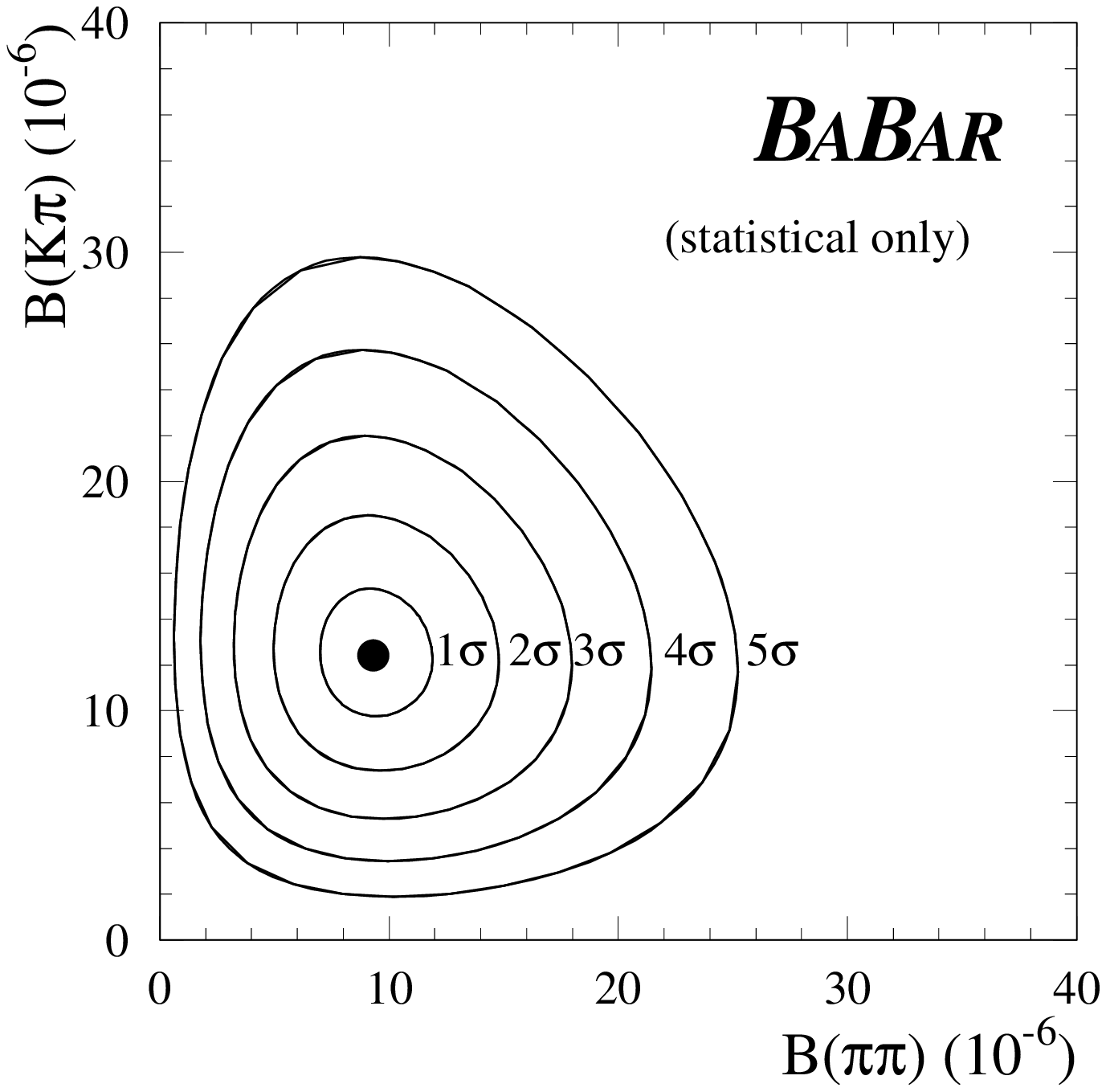}
\vspace{-0.6cm}
\caption{The central value (filled circle) for $\BR(\Bz\to \pip\pim)$ and 
$\BR(\Bz\to \Kp\pim)$ along with the $n\sigma$ statistical contour curves for 
the global likelihood fit, where $n\sigma$ corresponds to a change of $n^2$ in 
$-2\log{\cal L}$.}
\label{fig:sig}
\end{center}

\end{figure}


\section{Summary}

PEP-II and \BaBar\  had a beautiful start.
In their first year of data taking, they have
reached about 85\% of the design peak luminosity,
exceeded the design daily luminosity 
and accumulated almost 20~fb$^{-1}$ on tape. 

The first physics results, based on about 8 fb$^{-1}$,
have been produced.
This talk reviews the first results obtained by \BaBar\  in the measurement of 
$\sin2\beta$
and in the studies of B $\to$ charmonium and B $\to$ two body
charmless decays.

The results for the data collected in the entire first run of data taking 
(about 24 fb$^{-1}$) is expected to be available in Spring 2001. 


\section*{Acknowledgments}
The work presented in this report was supported by the Department of
Energy contract DE-AC03-76SF00515.

\end{document}